\documentclass[aps,prd,twocolumns,preprintnumbers,showpacs,amsmath,amssymb]
{revtex4}

\usepackage{graphicx}
\usepackage{dcolumn}
\usepackage{bm}

\begin{document}

\title{COMMENT TO: "A NOTE ON FAILURE OF THE LADDER APPROXIMATION TO QCD"

[PHYS. LETT. B 640 (2006) 196]}

\author{V. Gogokhia}
\email[]{gogohia@rmki.kfki.hu}

\affiliation{HAS, CRIP, RMKI, Depart. Theor. Phys., Budapest 114,
P.O.B. 49, H-1525, Hungary}

\date{\today}
\begin{abstract}
In the paper [Hong-Shi Zong, Wei-Min Sun, Phys. Lett. B 640 (2006)
196], the authors claim that our proof of the inconsistency of the
ladder approximation to QCD [Phys. Lett. B 611 (2005) 129] was
incorrect. However, their claim is based on a derivation which
contains a rough mathematical mistake, namely the unjustified
change of variables in the divergent (though regularized)
integrals. In this comment I will show this explicitly, so our
conclusion that the ladder approximation to QCD is inconsistent
remains, of course, correct.
\end{abstract}



\maketitle



Unfortunately for me, only these days by accident I have read the
paper \cite{1} in which the authors tried to refute my conclusion
on the inconsistency of the ladder approximation (LA) to QCD made
in my paper \cite{2}. However, unfortunately for them, their
derivation of Eq. (17) contains a rough mathematical mistake,
namely the unjustified change of variables in the divergent
(though regularized) integrals. In their derivation they have made
a replacement of the variable $l-k$ by $l'$ and thus $p-l = p-k -
l'$ in the third line of Eq. (17) in order to make use of the
quark DSE (1) of their paper. In other words, they naively equate

\begin{equation}
\int {d^4l \over {(2\pi)^4}} \gamma_\alpha S(l-k) \gamma_\beta
D_{\alpha \beta}(p-l) =  \int {d^4l' \over {(2\pi)^4}}
\gamma_\alpha S(l') \gamma_\beta D_{\alpha \beta}(p-k-l').
\end{equation}
Let us note that here and below the color group factors and
generators, as well as the dependence on the coupling constant
squared will be omitted, for simplicity, and will be restored in
the final expression only. However, such a replacement is strictly
forbidden in the divergent (though regularized) integrals (these
integrals are indeed divergent due to asymptotic freedom in QCD
\cite{3}).  How to carry out a replacement of variables in the
divergent integrals is explained almost in all the text-books on
gauge theories. I would recommend the authors of Ref. \cite{1} to
look, for example at Refs. \cite{4,5}, namely the chapters where
the vacuum polarization tensor is explicitly calculated in lower
order of the perturbation theory (PT). Here the Green's functions
$S$ and $D$ are not present by their free PT counterparts, so the
direct correct calculation in accordance with Refs. \cite{4,5} is
impossible. However, if one goes to use such a kind of replacement
of the loop variable, then the uncertainty of this procedure in
the divergent integrals should be explicitly taken into account.
So the correct relation (1) is

\begin{equation}
\int {d^4l \over {(2\pi)^4}} \gamma_\alpha S(l-k) \gamma_\beta
D_{\alpha \beta}(p-l) = \int {d^4l' \over {(2\pi)^4}}
\gamma_\alpha S(l') \gamma_\beta D_{\alpha \beta}(p-k-l') +
F(p,k),
\end{equation}
where the completely arbitrary function $F(p,k)$ just reflects the
above-mentioned uncertainty. Then their wrong Eq. (18) finally
becomes

\begin{equation}
C_A T ^a \int {d^4l \over {(2\pi)^4}} \gamma_\alpha (S(l-k) -S(l))
\gamma_\beta D_{\alpha \beta}(p-l) = C_F T^a F(p,k),
\end{equation}
after using of the WT identity (7) in both sides of their Eq.
(17). Here the dependence on the coupling constant squared is
omitted, for simplicity. Thus one obtains the result which leads
nowhere, i.e., some arbitrary function is expressed as the
difference between the two divergent unknown integrals. This
derivation has been done during my investigation. However, I did
not even mention it in my paper \cite{2} due to its triviality.
In my paper \cite{2} I decided not to use such a replacement of
the loop variable, but to use only the identity $S(p)S^{-1}(p) =1$
and its derivative.

{\bf That is why my conclusion of the inconsistency of the LA to
QCD made in Ref. \cite{2} is correct and the derivation in Ref.
\cite{1} is wrong. So my second main conclusion that all the
results in QCD based on the LA should be reconsidered remains
valid as well}.

Moreover, beside this mathematical mistake they have also made a
conceptual mistake stating that the WT identity should not be used
because it does not hold in the LA to QCD. The role of the
corresponding identities in any truncation/approximation schemes
has been already discussed in the Introduction to my paper
\cite{2}. Let me only note here that the WT identity is not
connected with any specific approximation, for example the PT. In
other words, it cannot be approximated, it is either identity or
it is not an identity at all. Correctly neglecting ghosts in the
LA (otherwise it will not be the LA), the ST identity becomes the
WT identity. However, it remains exact and all the approximations
correctly formulated should satisfy it. The wrong derivations, of
course, will not satisfy it, as it was just happened with the
authors of Ref. \cite{1}.

The reason of the inconsistency of the LA to QCD is that the color
charges interaction is always present in QCD, and this
approximation is not able to correctly take it into account. Such
kind of the interaction is absent in QED, and thus the LA works
there (in more detail the comparison between QCD and QED is
discussed in our paper \cite{2}).

\end{document}